\documentclass[prl,nobibnotes,nofootinbib, twocolumn,superscriptaddress,amsmath,twocolumn,amssymb]{revtex4}
\usepackage{color}
\usepackage{amsmath}
\usepackage{graphicx}
\usepackage[unicode=true, bookmarks=false, breaklinks=true,pdfborder={0.1 0.1 1},colorlinks=false] {hyperref}

\makeatletter

\DeclareTextSymbolDefault{\textquotedbl}{T1}

\@ifundefined{textcolor}{}
{%
 \definecolor{BLACK}{gray}{0}
 \definecolor{WHITE}{gray}{1}
 \definecolor{RED}{rgb}{1,0,0}
 \definecolor{GREEN}{rgb}{0,1,0}
 \definecolor{BLUE}{rgb}{0,0,1}
 \definecolor{CYAN}{cmyk}{1,0,0,0}
 \definecolor{MAGENTA}{cmyk}{0,1,0,0}
 \definecolor{YELLOW}{cmyk}{0,0,1,0}
}

\global\long\def\ket#1{\left|#1\right\rangle }%
\global\long\def\bra#1{\left\langle #1\right|}%
\global\long\def\creatop#1{\hat{a}_{#1}^{\dagger}}%
\global\long\def\anniop#1{\hat{a}_{#1}}%
\global\long\def\myi{\mathrm{i}}%
\global\long\def\mye{\mathrm{e}}%
\global\long\def\secref#1{sec.~\ref{=00003D00003D00003D00003D00003D00003D00003D00003D00003D00003D00003D00003D00003D00003D00003D00003D00003D00003D0000231}}%
\global\long\def\figref#1{Fig.~\ref{=00003D00003D00003D00003D00003D00003D00003D00003D00003D00003D00003D00003D00003D00003D00003D00003D00003D00003D0000231}}%
\makeatother

\begin{document}
\global\long\def\ket#1{\left|#1\right\rangle }%

\global\long\def\bra#1{\left\langle #1\right|}%

\global\long\def\bigket#1{\bigg|#1\bigg\rangle}%

\global\long\def\bigbra#1{\bigg\langle#1\bigg|}%

\global\long\def\n{\frac{1}{\sqrt{2}}}%

\global\long\def\mye{\text{e}}%

\global\long\def\myi{\text{i}}%

\global\long\def\creatop#1{\hat{a}_{#1}^{\dagger}}%

\global\long\def\anniop#1{\hat{a}_{#1}}%

$
\global\long\def\units#1{\ensuremath{\,\mathrm{#1}}}%
$
\title{How to administer an antidote to Schr\"{o}dinger's cat}
\author{Juan-Rafael \'{A}lvarez}
\author{Mark IJspeert}
\author{Oliver Barter}
\author{Ben Yuen}
\altaffiliation{now at Metamaterials Research Centre, University of Birmingham, Edgbaston, Birmingham, B15 2TT, UK}
\author{Thomas D. Barrett}
\altaffiliation{now at InstaDeep, London, UK.}
\author{Dustin Stuart}
\author{Jerome Dilley}
\altaffiliation{now at Google.}
\author{Annemarie Holleczek}
\altaffiliation{now at Robert Bosch GmbH, Postfach 13 55, 74003 Heilbronn, Germany}
\author{Axel Kuhn}
\email{axel.kuhn@physics.ox.ac.uk}

\affiliation{University of Oxford, Clarendon Laboratory, Parks Road, Oxford OX1
3PU, UK}
\date{\today}
\begin{abstract}
In his 1935 Gedankenexperiment, Erwin Schr\"{o}dinger imagined a poisonous substance which has a 50\% probability of being released, based on the decay of a radioactive atom. As such, the life of the cat and the state of the poison
become entangled, and the fate of the cat is determined upon opening
the box. We present an experimental technique that keeps the cat alive on any account. This method relies on the time-resolved Hong-Ou-Mandel
effect: two long, identical photons impinging on a beam splitter
always bunch in either of the outputs. Interpreting the first photon
detection as the state of the poison, the second photon is identified 
as the state of the cat. Even after the collapse of the first photon's state, we show their fates are intertwined through quantum interference. We demonstrate this by a sudden phase change between the inputs, administered conditionally on the outcome of the first detection, which steers the second photon to a pre-defined output and ensures that the cat is always observed alive.
\end{abstract}
\maketitle

\section{Introduction}

One of the most intriguing principles of quantum mechanics is that of superposition, which states that a quantum system, before being measured, can be interpreted to be
in two simultaneous states at once. In his 1935
Gedankenexperiment \citep{schroedingerGegenwaertigeSituationQuantenmechanik1935},
Erwin Schr\"{o}dinger illustrated the paradoxical nature of superposition
by depicting a cat in a box whose state (dead or alive) is entangled
with a vicious device releasing a poisonous substance upon a 50\%
probable radioactive decay. Only upon opening the box for the first
time, is it possible to determine the state of the combined
system of the cat and the radioactive device; with no decay together with an intact vile and a cat observed alive only half of the time.

The randomness of quantum measurements introduces a fundamental distinction
with respect to classical measurements. The process of measurement
is central to many open theoretical questions \citep{adlam_2021}, where the measurement induces
apparent contradictions between the predictions of quantum mechanics
and the appearance of sharp measurement outcomes \citep{landsmanMeasurementProblem2017}. In addition, many systems and applications involve a form of quantum control that relies on the quantum nature of measurements \citep{lloydQuantumControllersQuantum1997}.

This poses the question: can a partial measurement of a quantum system be made such that it triggers a sequence of events that coerces the remainder of the system into a desired state? In the example of Schr\"{o}dinger's cat, such an ability would ensure that the cat is always observed alive, whether or not the poison has been released. Such an approach would normally be implemented in the form of a feedback loop, by which a device obtains information about the trajectory of a physical
system in order to modify it in real time \citep{zhangQuantumFeedbackTheory2017,habibQuantumFeedbackControl2002}.

In quantum systems, feedback can be of two types: the first, measurement-based
quantum feedback, occurs when a measurement outcome defines a subsequent action on the original system. The second, coherent quantum feedback, involves
no measurements but provides control using coherent interaction with
an auxiliary quantum system.

Both feedback types exist in a range of applications: the
generation of amplitude squeezed states in a semiconductor laser \citep{yamamotoAmplitudeSqueezingSemiconductor1986},
the improvement of single-shot phase measurements in quantum metrology
\citep{armenAdaptiveHomodyneMeasurement2002a}, the stabilisation
of a combined atom-cavity quantum state \citep{smithCaptureReleaseConditional2002a},
and the preparation and stabilisation of Fock states in a high-Q microwave
cavity with weak measurements \citep{sayrinRealtimeQuantumFeedback2011a,zhouFieldLockedFock2012a}.

In this work, we take advantage of a measurement-based feedback protocol to deliver a photonic state with a desired property: that of always exiting through the output of a beam splitter of the experimentalist's choosing. Specifically, the feedback process is applied in a time-resolved two-photon quantum interference experiment  \citep{legeroTimeresolvedTwophotonQuantum2003} in which the time span between the first and second photon
detection is long enough ($\sim500\units{ns}$) for a phase
change to be applied on the second photon. This  allows us to alternate between the
behaviours of bosonic and fermionic interference to control the routing
of a single photon as desired. 

\begin{figure*}
\begin{centering}
\includegraphics[width=2\columnwidth]{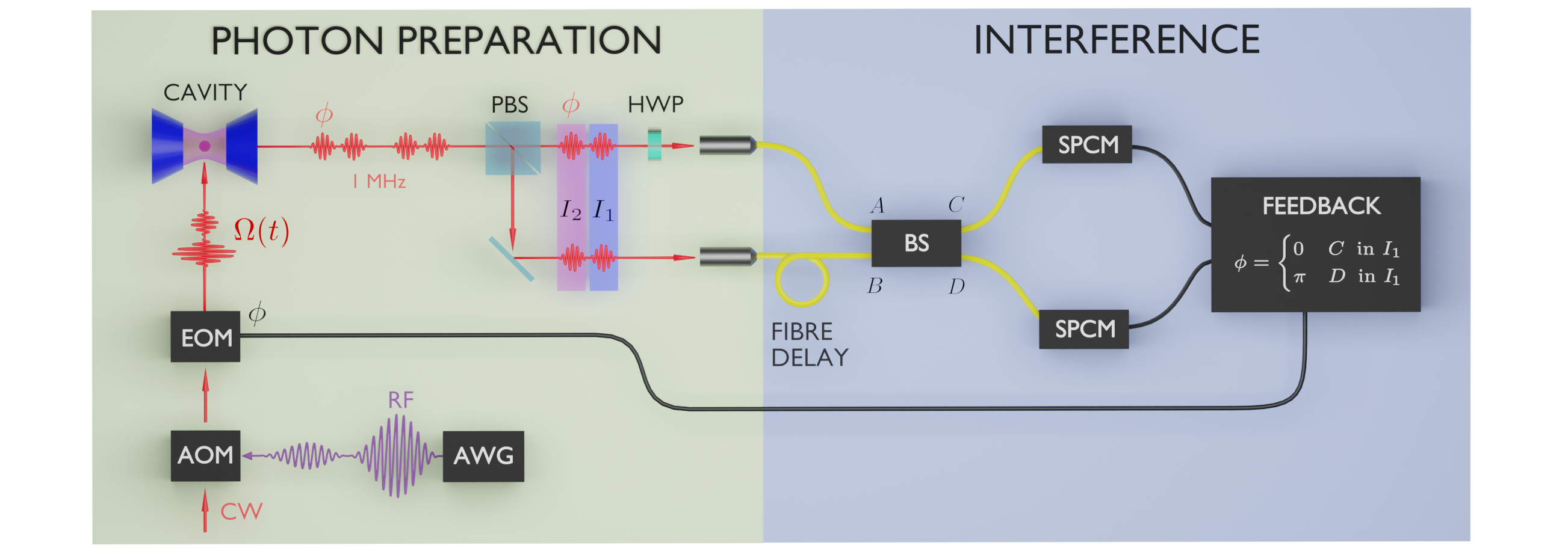}
\par\end{centering}
\caption{Experimental arrangement:
Two long, double-hump photons interfere in a beam splitter. Photons
are emitted at a repetition rate of $1\units{MHz}$ from an atomic source driven by a laser pulse controlled by an acousto-optic modulator (AOM), whose
phase is changed using an electro-optic modulator (EOM). The
AOM pulses are generated by an arbitrary waveform generator (AWG). A
delay line with an optical path length of $300\units{m}$ ensures the simultaneous arrival of two sequentially emitted photons. The emitted photons have such long wave packets,
that the first time bin of both photons interferes before the relative
phase $\phi$ of the second photon in the second time bin is set.
Furthermore, the emitted photons have random polarisations. Using
a polarising beam splitter (PBS), photons are routed into random paths,
and two photons simultaneously impinge on the input ports $A$ and
$B$ of a beam splitter (BS) in 25\% of all  possible cases. The
relative polarisation between both photons is changed using a
half wave plate (HWP). Photons then interfere in the BS and exit through
the ports $C$ and $D$, upon which measurements are performed using single
photon counting modules (SPCMs). The outcome of a measurement in the
first time bin is used to perform feedback on the value of the
phase, $\phi$ during the second half of the driving pulse. \label{fig:Idea_Setup}}
\end{figure*}

\section{Theory}

Let us consider two photons arriving simultaneously at the input ports
$A$ and $B$ of a beam splitter (BS), as illustrated in the Interference box of Fig. \ref{fig:Idea_Setup}. The joint probability of detecting
two photons at times $t_{0}$ and $t_{0}+\tau$ at detectors placed
at outputs $C$ and $D$ of the BS, respectively, can be written as
 \citep{legeroTimeresolvedTwophotonQuantum2003,legero06}
\begin{equation}
\begin{aligned}P_{\text{joint }}\left(t_{0},\tau\right)=\left|\hat{E}_{D}^{+}\left(t_{0}+\tau\right)\hat{E}_{C}^{+}\left(t_{0}\right)\hat{a}_{A}^{\dagger}\hat{a}_{B}^{\dagger}\ket{\text{0}_{A}0_{B}}\right|^{2}\end{aligned}
,\label{eq:Pjoint}
\end{equation}
where $\hat{a}_{A}^{\dagger}$ and $\hat{a}_{B}^{\dagger}$ correspond to the creation operators
of a photon in the ports $A$ and $B$ ($\hat{a}_{A}$ and $\hat{a}_{B}$ would correspond
to the annihilation operators), $\hat{E}_{C}^{+} + \hat{E}_{C}^{-}$  and $\hat{E}_{D}^{+} + \hat{E}_{D}^{-}$ are the electric field operators at the output ports of the beam splitter, and $\ket{0_{A}0_{B}}$
corresponds to the vacuum state on the input ports $A$ and $B$.
In this context, $\tau$ can be positive or negative, and $\tau<0$ corresponds
to the detector in port $D$ clicking before that in port $C$. We
emphasise that $\tau$ is a detection time difference giving rise
to a time-resolved Hong-Ou-Mandel (HOM) signal \citep{legeroTimeresolvedTwophotonQuantum2003,legero06},
and must not be confused with the photon arrival time difference $\left(\Delta t\right)$
used in many other HOM experiments \citep{hongMeasurementSubpicosecondTime1987a}.
Here, we are only interested in the case for which the two photon wavepackets
arrive simultaneously ($\Delta t=0$). 

The electric field operators can be written as the sum of spatio-temporal
functions in distinct modes $k$, $\zeta_{k}\left(t\right)=\epsilon_{k}\left(t\right)\exp\left(-i\phi_{k}\left(t\right)\right)$:
\begin{equation}
\hat{E}^{+}(t)=\sum_{k}\zeta_{k}(t)\hat{a}_{k}\text{, \quad }\hat{E}^{-}(t)=\sum_{k}\zeta_{k}^{*}(t)\hat{a}_{k}^{\dagger},
\end{equation}
where $\epsilon_{k}\left(t\right)$ corresponds to the photon amplitude
in mode $k$ and $\phi_{k}\left(t\right)$ to its phase.

Since the absolute photon detection times are irrelevant, we calculate the joint detection probability as a function of $\tau$ only:
\begin{equation}
P_{\text{joint}}\left(\tau\right)=\int_{-\infty}^{\infty}dt_{0}P_{\text{joint}}\left(t_{0},\tau\right).\label{eq:PjointOneVar}
\end{equation}

For two linearly polarised photons with a relative polarisation angle
$\theta$, Eq. \ref{eq:Pjoint} can be written as \citep{legero06}
\begin{equation}
\begin{aligned}P_{\text{joint}}\left(t_{0},\tau\right)=P_{\text{joint}}^{(HV)}\left(t_{0},\tau\right)-\cos^{2}\theta\text{ }F\left(t_{0},\tau\right)\end{aligned}
,\label{eq:AnyPhotons}
\end{equation}
where 
\begin{multline}
P_{\text{joint}}^{(HV)}\left(t_{0},\tau\right)=\frac{1}{4}\left(\left|\epsilon_{A}\left(t_{0}\right)\epsilon_{B}\left(t_{0}+\tau\right)\right|^{2}+\right.\\
\left.\left|\epsilon_{A}\left(t_{0}+\tau\right)\epsilon_{B}\left(t_{0}\right)\right|^{2}\right)\label{eq:PhaseIndep}
\end{multline}
and
\begin{multline}
F\left(t_{0},\tau\right)=\frac{\epsilon_{A}\left(t_{0}\right)\epsilon_{B}\left(t_{0}+\tau\right)\epsilon_{A}\left(t_{0}+\tau\right)\epsilon_{B}\left(t_{0}\right)}{2}\times\\
\cos\left(\phi_{A}\left(t_{0}\right)-\phi_{A}\left(t_{0}+\tau\right)+\phi_{B}\left(t_{0}+\tau\right)-\phi_{B}\left(t_{0}\right)\right).
\end{multline}

When both input photons have orthogonal polarisations $\left(\theta=\pi/2\right)$,
$P_{\text{joint}}\left(\tau\right)$ can be written as the convolution
of their spatio-temporal squared amplitudes, $P_{\text{joint}}\left(\tau\right)=\frac{1}{2}\left(\left|\varepsilon_{A}\right|^{2}*\left|\varepsilon_{B}\right|^{2}\right)(\tau),$
rendering the choice of $\phi_{A}$ and $\phi_{B}$ irrelevant. This
is in sharp contrast to the case for which both input photons have parallel
polarisations $\left(\theta=0\right)$, where $\phi_{A}$ and $\phi_{B}$
become relevant. The expected behaviour of $P_{\text{joint}}\left(\tau\right)$
for the particular case of two photons of perpendicular polarisation
with the temporal shape $\epsilon\left(t\right)=\sin^{2}\left(2\pi t/\delta t\right)$,
is shown as a dashed curve in Fig. \ref{fig:feedbackResults}(a) for
$\delta t=401\units{ns}$. The three peaks that are shown in Fig.
\ref{fig:feedbackResults}(a) correspond to the detection of coincidences
in detectors $C$ and $D$ delayed by varying time differences $\tau\in\left[-401,401\right]\text{ ns}$.

We subdivide the overall duration of the photons into two distinct
intervals (bins) of equal length labeled $I_{1}$ (early) and $I_{2}$
(late). Additionally, let us assume that, for the photon arriving
through $A$, $\phi_{A}\left(t\right)=0$ for all times, and that,
for the photon arriving through $B$, $\phi_{B}\left(t\right)=0$
for $t\in I_{1}$, and $\phi_{B}\left(t\right)=\phi$ for $t\in I_{2}$. In the Schr\"{o}dinger picture \citep{legero06}, the
state entering the BS is given by 
\begin{equation}
\ket{\Psi_{\text{in}}}=\frac{1}{2}\left(\creatop{A1}+\creatop{A2}\right)\left(\creatop{B1}+\mye^{\myi\phi}\creatop{B2}\right)\ket 0.\label{eq:BothPhotonsInputPhase}
\end{equation}
Here, $\hat{a}_{Aj}^{\dagger}$ and $\hat{a}_{Bj}^{\dagger}$ correspond
to the photon creation operators in ports $A$ and $B$ in the time
intervals $I_{j}$, with $j\in\{1,2\}$, and $\ket 0$ is the vacuum
state in the basis of all the temporal and beam splitter input paths:
$\ket 0=\ket{0_{A1}0_{A2}0_{B1}0_{B2}}$. For the cases $\phi=0$
and $\phi=\pi$, the theoretical predictions of $P_{\text{joint}}\left(\tau\right)$
are shown as dashed curves in Figs. \ref{fig:feedbackResults}(b)
and (c). For the case of Fig. \ref{fig:feedbackResults}(b), both
input photons are identical and feature photon bunching in the
output detectors. For this reason, the expected behaviour would not
include any coincidences between detectors $C$ and $D$. In contrast,
in Fig. \ref{fig:feedbackResults}(c) the photon entering through
port $B$ acquired a $\pi$-phase change from $I_{1}$ to $I_{2}$. This
is the only difference between the two photons. Therefore, if both
photons are detected during either $I_{1}$ or $I_{2}$, the photons
are indistinguishable and no correlations are found with $\tau\simeq0$.
However, if the photons are detected in different intervals, the change
in phase drives them to different outputs, resulting in a coincidence
probability that is twice as large as the orthogonal polarisation
case for $\tau=\pm201\units{ns}$.

Figs. \ref{fig:feedbackResults}(b) and (c) exhibit a form that corresponds
to a different type of interference between the two photons depending
on the value of $\phi$. Bosonic ($\phi=0$) and fermionic interference
($\phi=\pi$) is found with photons either bunching in the same outputs
or avoiding each other and giving rise to coincidences, respectively.
To see this, consider the following argument, using the Schr\"{o}dinger
picture: the operators after the beam splitter in the output channels
$C$ and $D$ in $I_{j}$ are linked to those before it by the standard
unitary relation $\creatop{Aj,Bj}=\left(\creatop{Cj}\pm\creatop{Dj}\right)/\sqrt{2}.$
Therefore, after the beam splitter, Eq. \ref{eq:BothPhotonsInputPhase}
results in:
\begin{multline}
\ket{\Psi_{\text{out}}}=\frac{1}{2\sqrt{2}}\left(\left(\creatop{C1}\creatop{C1}+\mye^{\myi\phi}\creatop{C2}\creatop{C2}\right.\right.\\
-\left.\creatop{D1}\creatop{D1}-\mye^{\myi\phi}\creatop{D2}\creatop{D2}\right)\\
\left.+\left(\mye^{\myi\phi}+1\right)\left(\creatop{C1}\creatop{C2}-\creatop{D1}\creatop{D2}\right)\right.\\
\left.+\left(\mye^{\myi\phi}-1\right)\left(\creatop{C2}\creatop{D1}-\creatop{C1}\creatop{D2}\right)\right)\ket{0}.\label{eq:equ3}
\end{multline}
Here, $\ket{0}$ is represented in the basis of all the
output ports and temporal modes: $\ket{0}=\ket{0_{C1}0_{C2}0_{D1}0_{D2}}$.
The first four terms account for both photons arriving at the same
detector in the same time bin. These terms lead to the standard boson
bunching reported in the canonical HOM effect \citep{hongMeasurementSubpicosecondTime1987a}.
The last four terms correspond to cross-correlations between detectors
and/or time bins. Unless the single photon detectors are number resolving,
it is not possible to measure outcomes where both detections occur
at the same detector in the same time bin. By eliminating the terms
of the form $\creatop{Xi}\creatop{Xi}$ and rearranging the terms,
the observable sub-state from Eq. \ref{eq:equ3} becomes 

\begin{multline}
\ket{\tilde{\Psi}_{\text{out}}}=\frac{1}{2\sqrt{2}}\left(\hat{a}_{C1}^{\dagger}\left(\left(e^{i\phi}+1\right)\hat{a}_{C2}^{\dagger}-\left(e^{i\phi}-1\right)\hat{a}_{D2}^{\dagger}\right)\right.\\
\left.+\hat{a}_{D1}^{\dagger}\left(\left(e^{i\phi}-1\right)\hat{a}_{C2}^{\dagger}-\left(e^{i\phi}+1\right)\hat{a}_{D2}^{\dagger}\right)\right)\ket{0}.\label{eq:equ4-1}
\end{multline}

In this sub-state, the phase appears explicitly. By setting $\phi=0$,
Eq. \ref{eq:equ4-1} reads

\begin{equation}
\ket{\tilde{\Psi}_{\text{out}}^{\left(\phi=0\right)}}=\frac{1}{\sqrt{2}}\left(\creatop{C1}\creatop{C2}-\creatop{D1}\creatop{D2}\right)\ket{0}.\label{eq:Bosonic}
\end{equation}
This state again leads to the canonical HOM effect, yet
across the time bins this time. Therefore, both photons arrive at in the same detector,
but in separate time intervals. In contrast, setting $\phi=\pi$,
Eq. \ref{eq:equ4-1} now gives 

\begin{equation}
\ket{\tilde{\Psi}_{\text{out}}^{\left(\phi=\pi\right)}}=\frac{1}{2\sqrt{2}}\left(\creatop{C1}\creatop{D2}-\creatop{C2}\creatop{D1}\right)\ket{0},\label{eq:Fermionic}
\end{equation}
which is strikingly different to Eq. \ref{eq:Bosonic}, as both photons
exhibit fermionic behaviour by arriving at different detectors. 

Switching between the bosonic and fermionic behaviour
by choice of $\phi$ allows one to put the system into a pre-defined
quantum state conditioned on the random outcome of the first measurement.
This is implemented in the form of a feedback mechanism, where a photon
detection at either detector in $I_{1}$ reveals the required change of $\phi$
to steer the remaining photon in $I_{2}$ to a specific detector.
For instance, if detector $D$ clicks in $I_{1}$, the second detection
would normally occur in detector $D$, due to the canonical HOM effect.
However, as the second halves of the photons have not reached the
input ports at the moment of detection, we can instantly change the
phase between $I_{1}$ and $I_{2}$ for one of the incoming photons
to change the port in which the second detection occurs. By choosing
$\phi=\pi$ and projecting the output state in Eq. \ref{eq:Fermionic}
onto the measured state $\ket{0_{C1}1_{D1}}=\hat{a}_{D1}^{\dagger}\ket{0_{C1}0_{D1}}$
(involving only the first time bin, since the second one has not yet
occurred), the state reduces to

\begin{equation}
\bigbra{0_{C1}0_{D1}}\hat{a}_{D1}\bigket{\tilde{\Psi}_{\text{out}}^{\left(\phi=\pi\right)}}=\frac{1}{\sqrt{2}}\hat{a}_{C2}^{\dagger}\ket{0_{C2}0_{D2}}.\label{eq:PhiPi_Feedback}
\end{equation}
This state clearly shows that, when a measurement occurs in $I_{2}$,
any click will be recorded in detector $C$. Analogously, if detector
$C$ clicks in $I_{1}$ and $\phi=0$ is chosen, both photons are
identical in their two halves and any detection in $I_{2}$ will
always occur at detector $C$. This reduces the quantum state to 

\begin{equation}
\bigbra{0_{C1}0_{D1}}\hat{a}_{C1}\bigket{\tilde{\Psi}_{\text{out}}^{\left(\phi=0\right)}}=\frac{1}{\sqrt{2}}\hat{a}_{C2}^{\dagger}\ket{0_{C2}0_{D2}}.\label{eq:Phi0_Feedback}
\end{equation}
Again, when a measurement occurs in $I_{2}$, any click will be recorded
by detector $C$. Eqs. \ref{eq:PhiPi_Feedback} and \ref{eq:Phi0_Feedback}
imply that $\phi$ can be used as a parameter for feedback control, to steer the remaining photon to a desired output.

Returning to Eqs. \eqref{eq:PjointOneVar}
and \eqref{eq:AnyPhotons} with a value of $\phi$ conditional on the
measurement of a photon in $I_{1}$, we observe the
expected behaviour of $P_{\text{joint}}\left(\tau\right)$, as illustrated
in Fig. \ref{fig:feedbackResults}(d). Fig. \ref{fig:feedbackResults}(d)
shows increased coincidences with $\tau<0$, for which a detection in
$D$ occurs before a detection in $C$. There are significantly less coincidences with
$\tau>0$, as this would correspond to detector $C$ firing first
and detector $D$ second, which cannot occur with an active feedback that always prompts the second detection in $C$. Fig. \ref{fig:feedbackResults} shows the subset of probabilities for cross-detector correlations only. For this reason, the value $\int P_{\text{joint}}\left(\tau\right)d\tau$ is bounded by $1/2$. The theoretical curves for same-detector correlations are shown in the Supplementary Material.

\section{Experimental methods}

\begin{figure}
\begin{centering}
\includegraphics[width=1\columnwidth]{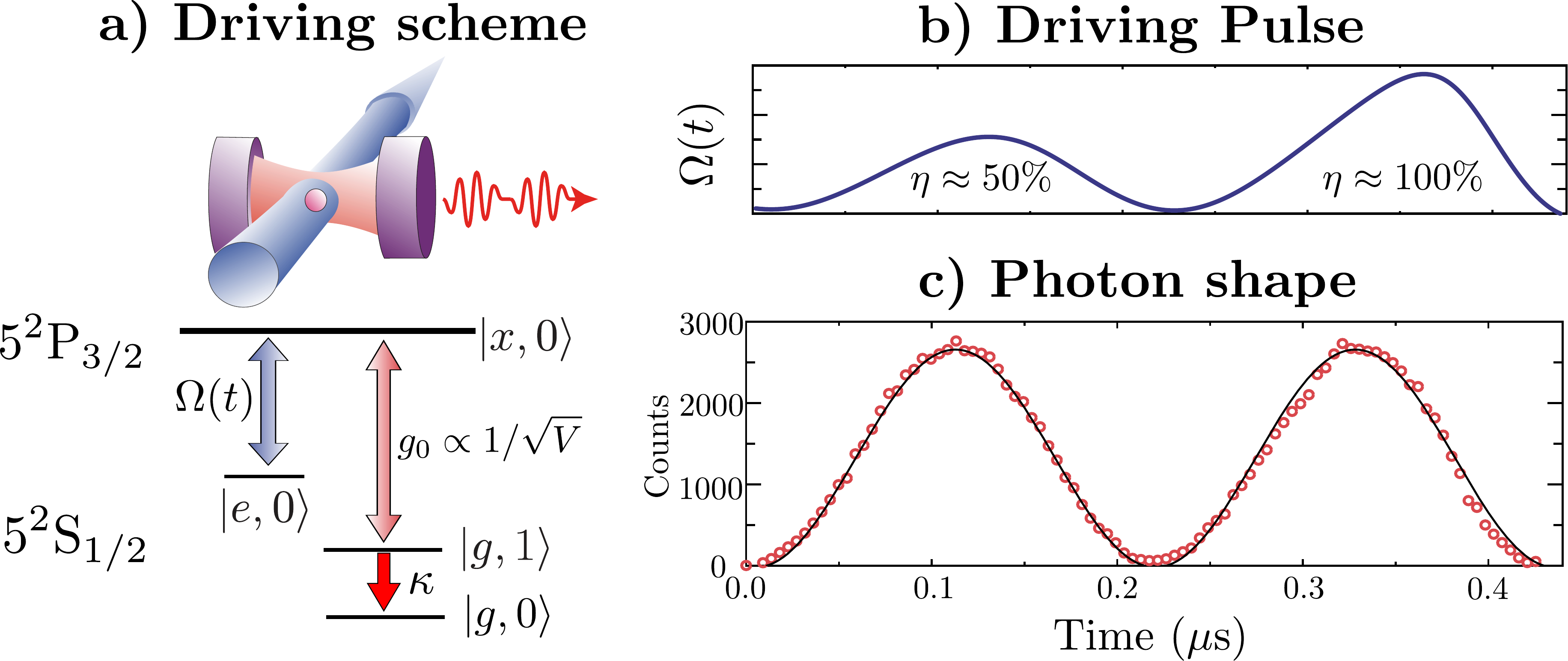}
\par\end{centering}
\caption{\textcolor{black}{\label{fig:doublehumpedphotons}}\textcolor{black}{\emph{
}}\textcolor{black}{Photon production scheme for the generation of
double-hump photons. Fig (a) shows the driving scheme used to generate
single photons for a single $^{87}\mathrm{Rb}$ atom inside of an
optical cavity, following a coherent STIRAP process between the $\protect\ket{F=1}\equiv\protect\ket e$
and $\protect\ket{F=2}\equiv\protect\ket g$ ground states of $^{87}\mathrm{Rb}$
via the virtually excited state $\protect\ket{F'=3}\equiv\protect\ket x$
\citep{nisbet-jonesHighlyEfficientSource2011}. (b) shows the pre-calculated
\citep{vasilevSinglePhotonsMadetomeasure2010} pulse necessary for
the photon shape to follow the form $\sin^{2}(2\pi\tau)$ with $0<\tau<1$,
where $\tau=t/450\units{ns}$, as can be seen theoretically (solid
line) and experimentally (dots) in (c). Our cavity has a decay rate
of $\kappa=2\pi\times12\units{MHz}$ and coupling strength of $g=2\pi\times15\units{MHz}$.}}
\end{figure}

To demonstrate this phenomenon experimentally, we generate photons
in an atom-cavity system using a standard V-STIRAP scheme \citep{kuhnCavitybasedSinglephotonSources2010}
coupling the hyperfine levels of the $D_{2}$ line of $^{87}\text{Rb}$.
Specifically, the $\ket e=\ket{5^{2}\text{S}_{1/2},F=1}$ and $\ket g=\ket{5^{2}\text{S}_{1/2},F=2}$
ground states are coupled in a $\Lambda$-scheme to the excited state
$\ket x=\ket{5^{2}\text{P}_{3/2},F^{\prime}=3}$ using a driving laser
$\Omega(t)$, as illustrated in Fig. \ref{fig:doublehumpedphotons}.
The cavity has a decay rate of $\kappa=2\pi\times12\units{MHz}$ and
a maximum coupling strength of $g_{0}=2\pi\times15\units{MHz}$. With
the system initially prepared in $\ket{e,0}$, the laser adiabatically
drives the system to $\ket{g,1}$, whereupon the photon is emitted
from the cavity mode, leaving the system in $\ket{g,0}$, decoupled
from further evolution.  Atoms are loaded into the cavity with an atomic fountain, from a magneto-optic trap (MOT) located directly under the cavity.

The spatio-temporal profile $\zeta\left(t\right)$ of the emitted
photons can be directly controlled by shaping the driving pulse \citep{nisbet-jonesHighlyEfficientSource2011}
using an acousto-optic modulator (AOM). For our purposes, it is
most crucial that the phase of the driving laser, which we control
via a separate electro-optic modulator (EOM), is directly mapped
to the phase of the emitted photon. Fig. \ref{fig:doublehumpedphotons}(b)
shows the driving pulse shape and Fig. \ref{fig:doublehumpedphotons}(c)
the resulting doubly-humped photon profile, evenly distributed
across two $225\units{ns}$ time-bins, for a total coherence time
of $\delta t=450\units{ns}$.\emph{ }

The experimental sequence for the production of photons is as follows:
photons are emitted at a repetition rate of $1\units{MHz}$, with $450\units{ns}$
of the cycle used for the production of a single photon. The remaining
$550\units{ns}$ of the cycle are used to optically repump the atom
to $\ket e$ to repeat the process. The photons impinge on a polarising
beam splitter (PBS), and are randomly routed into two paths, one of
which incorporates a fibre loop of $300\units{m}$ of optical path
length to induce a $1\units{\mu s}$ delay. This ensures that a pair
of subsequently emitted photons arrives simultaneously at the beam
splitter, which is the only case of interest. A half wave plate (HWP) sets the relative polarisation
of both photons.

Note from Fig. \ref{fig:doublehumpedphotons}(c), that the first halves
of the emitted photons span $67.5\units{m}$ ($\sim225\units{ns}$),
despite the optical path length between the cavity and the detectors
being only $1.5\units{m}$ or $301.5\units{m}$, depending on the
path taken. This means that the first time interval of the two photon
state is already being measured before the second half of the second
photon leaving the cavity (travelling along the shorter path) has
been fully generated. The length of the interfering photons is sufficient
for a feedback loop to alter the phase $\phi$ of the second half
of the photon under production, conditioned on the measurement outcome
within the first time interval (Fig. \ref{fig:Idea_Setup}.)

The feedback is implemented using a home-built circuit controller
and single photon counting modules (SPCMs) with a quantum efficiency
of 60-65\% and a resolution of $<300\units{ps}$ (Excelitas SPCM-AQRH-780-14-FC).
The total feedback loop latency, from the EOM via the cavity to the
SPCM, then back to the EOM via the controller circuitry is $97.0\pm0.2\units{ns}$. Therefore, a conditioned phase change cannot be realised in time
if a photon detection in $I_{2}$ occurs less than 97 ns after a detection
in $I_{1}$. The resulting error rate is limited to 0.2\% for a $\sin^{4}\left(t\right)$
photon intensity envelope. All photon counts are recorded with 81ps
accuracy using a qutools quTAU time-to-digital converter (TDC). Detections
within the $450\units{ns}$ photon window are further processed for
a dark count correction for each SPCM. A detailed description on the
error rates, time budgets, the implementation of the logical feedback
circuit and background correction is provided in the Supplemental
Material.

\section{Results}

\begin{figure*}
\begin{centering}
\includegraphics[width=2\columnwidth]{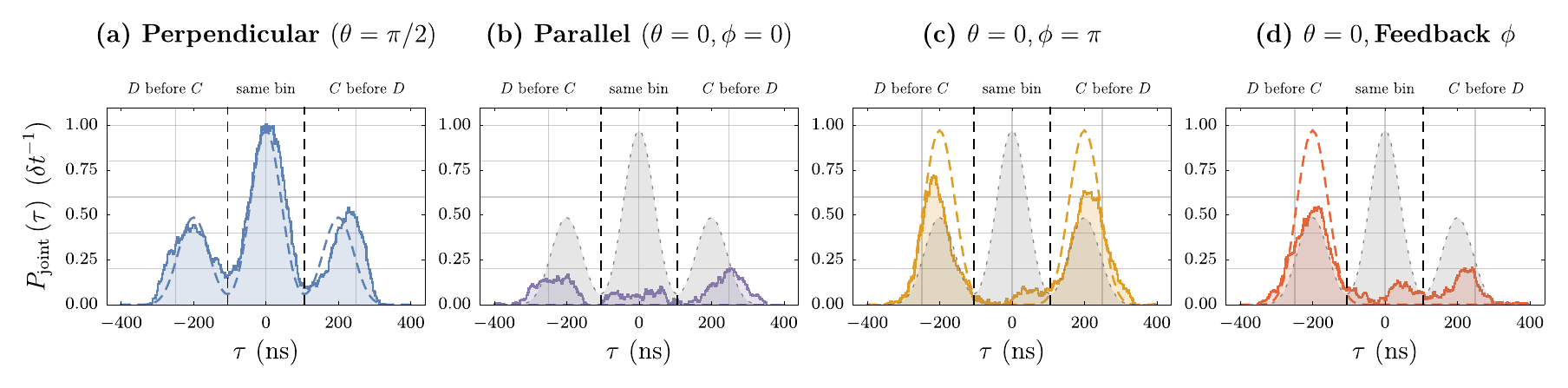} 
\par\end{centering}
\caption{Coincidence probability densities and sliding histograms (whose bin width is larger than its bin separation) showing the theoretical (dashed) and experimental (solid) values for $P_{\text{joint}}\left(\tau\right)$. This illustrates the  time-resolved HOM interference of two
photons under three different conditions. (a) shows the random routing
of photons with perpendicular polarisations, which serves as reference.
(b) shows the interference between indistinguishable photons ($\phi=0$)
with parallel polarisations\emph{, }resulting in almost no coincidences
between the two detectors. (c) shows the trace obtained for photons
with phase shifts of $\phi=\pi$ without feedback. (d) shows the asymmetric
pattern observed under feedback control, where the feedback has prevented
(most) correlations in the right satellite peak, steering the  second photon to a pre-defined output. This ensures that the cat is always observed alive. The data shown here has been corrected for correlations
involving background noise, as explained in the Supplemental Material.
\label{fig:feedbackResults} }
\end{figure*}

\begin{figure}
\centering{}\includegraphics[width=1\columnwidth]{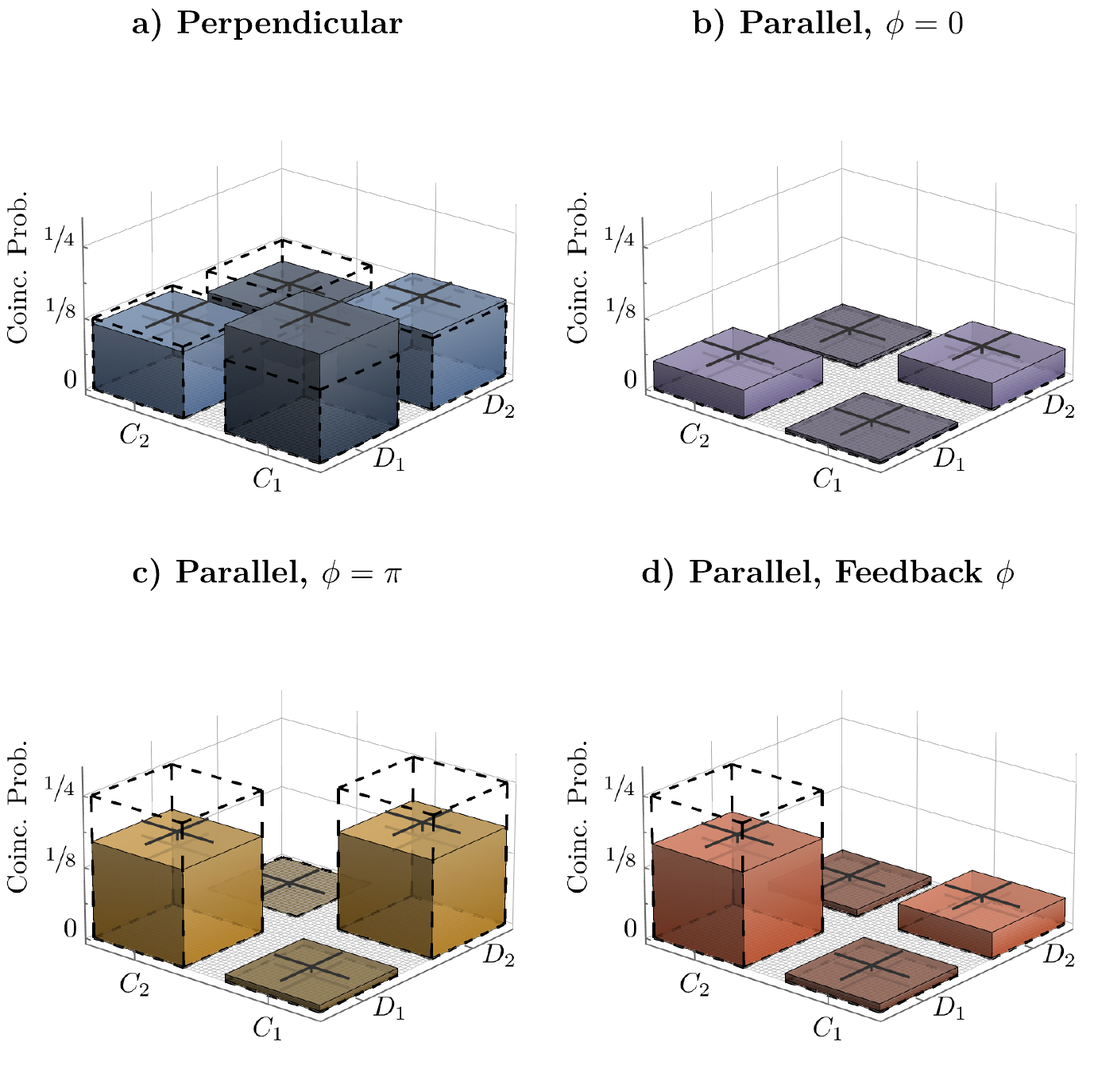}
\caption{The correlations from Fig. \ref{fig:feedbackResults} are now shown
according to the exact detector-time-bin detections. 
In (a), an increased number in the correlations between $C_1$ and $D_1$ with respect to $C_2$ and $D_2$ is  explained due to a larger number of photons in the first time bin, as shown in the Supplementary Material. In (d), we observe an increase in the correlations between $C_{2}$ and $D_{1}$ with respect to (a). This implies
that a detection in $D$ in $I_{1}$ has been used to steer the measurement
of a photon in $I_{2}$ to $C$, and serves as a demonstration that the second photon is sent to a pre-defined output state, i.e., the cat is always observed alive. A limitation in visibility is evident
when comparing the data to their theoretical values, shown dashed.
The data shown here has been corrected for correlations involving
background noise, as explained in the Supplementary Material. \label{fig:feedbackBar3D}}
\end{figure}

There are two different ways to look at the measured data. One has
been discussed previously and corresponds to Fig. \ref{fig:feedbackResults},
which shows the coincidence probability as a function of the time-difference
between two detections. However, our main interest is whether both
detections are registered within the same or opposite time intervals ($I_{1}$ and $I_{2}$). Therefore, the same data can be represented in a cross-correlation
diagram (Fig. \ref{fig:feedbackBar3D}) showing the four possible
values of the coincidence probability between detectors $C$ and $D$
firing in either $I_{1}$ or $I_{2}$. In contrast to Fig. 3, this
allows to further differentiate between coinciding detections in $I_{1}$
and $I_{2}$.

The cross detection probability of interfering photons with perpendicular polarisations is shown in Fig. \ref{fig:feedbackResults}(a). This gives
rise to the random routing of the simultaneously arriving photons,
such that the resulting time-resolved coincidence rate yields the
autocorrelation function of the photons' intensity profile. The cross-correlations
are depicted in Fig. \ref{fig:feedbackBar3D}(a). For the case of
photons with orthogonal polarisations, these are identical and theoretically
equal to $1/8$.

For photons with parallel polarisations, the cross correlation probability
is shown in Fig. \ref{fig:feedbackResults}(b) for $\phi=0$. This
measurement follows the predictions of photon bunching in the canonical
HOM effect, for which we expect no coincidences. Photons bunch in the
same output regardless of the actual detection time. Thus, the probability
of cross channel detections is expected to be zero for all time differences (Fig. \ref{fig:feedbackBar3D}(b)).

Fig. \ref{fig:feedbackResults}(c) shows the cross detection probability
for photons with parallel polarisations when $\phi=\pi$. In this
case, the photons antibunch (i.e., they are found in different output
channels if detected in different time intervals). The maximum probability
to find photon-photon correlations occurs at $\tau=\pm\delta t$,
where $\delta t$ is the length of the photons. This maximum probability is twice
the reference value of Fig. \ref{fig:feedbackResults}(a), which is
in accordance with the cross-channel detections shown in Fig \ref{fig:feedbackBar3D}(c).
Nonetheless, the likelihood of correlated photon detections in the same
time interval is close to its theoretical value of zero.

Finally, the cross detection probability shown in Fig. \ref{fig:feedbackResults}(d)
represents the case of active feedback on $\phi$, for which we expect
any photon recorded in $I_{2}$ be detected at $C$. Therefore,
the number of \textquotedbl$D\text{ before }C$\textquotedbl{} correlations
reaches a maximum while we barely see any \textquotedbl$C\text{ before }D$\textquotedbl{}
correlations, as shown in Fig. \ref{fig:feedbackBar3D}(d). Our experimental
results demonstrate that classical feedback control of a quantum excitation
spanning multiple systems (RF pulse driving AOM $\rightarrow$ driving laser $\rightarrow$
atom $\rightarrow$ cavity $\rightarrow$ quantum field modes) can be achieved, resulting in
a photonic state with the property of always exiting through the same
output of a beam splitter.

Some differences between theoretical expectations and experimental
results are visible in Figs. \ref{fig:feedbackResults} and \ref{fig:feedbackBar3D}.
These can be attributed to a partial loss of coherence or depolarisation
of the interfering photons, most evident from the presence of correlations
in the side lobes of Fig. \ref{fig:feedbackResults} (b). The mutual coherence between photons is often characterised by the HOM visibility, defined as 
\begin{equation}
V_{\text{HOM}}=1-\frac{N_{\parallel}}{N_{\perp}}
\end{equation}
where $N_{\parallel}$ and $N_{\bot}$ are the total number of correlations
observed for interfering photons with parallel and perpendicular polarisations,
respectively. \textcolor{black}{Using the results obtained with $\phi=0$
for $N_{\parallel}$ we find $V_{\text{HOM}}=0.78\pm0.04$. }However,
we measure a reduced visibility of $V_{\text{ref}}=0.61\pm0.04$ if we restrict our analysis to
those correlations with detections across time intervals $I_{1}$ and $I_{2}$.
This serves as a reference for all effects
discussed here, as these affect only the correlations across both time intervals.

The visibility under phase control upon switching between $\phi=0$ and $\phi=\pi$ reads 
\begin{equation}
V_{\phi}=\frac{N_{\pi}-N_{0}}{N_{\pi}+N_{0}},
\end{equation}
where $N_{0}$ and $N_{\pi}$ represent the coincidence counts between
pairs of photons with relative phase shifts of $\phi=0$ and $\pi$,
respectively. Again, we only count coincidences across $I_{1}$ and
$I_{2}$, and find $V_{\phi}=0.56\pm0.06$, in agreement
with $V_{\text{ref}}$ within error bars. This validates the robustness of the phase switch as no further loss of coherence is induced\footnote{$V_{\text{ref}}$ and $V_\phi$ are expected to yield the same value, assuming $N_\parallel=N_0$ and $N_\perp=\left(N_0+N_\pi\right)/2$ due to the random splitting of photons.}.

Since the feedback relies on a conditional phase switch, we quantify
its visibility by comparing coincidence counts between intervals where
a measurement in $D$ occurs before one in $C$ ($N_{D1C2}$),
and vice-versa ($N_{C1D2}$):

\begin{equation}
V_{\text{feed}}=\frac{N_{D1C2}-N_{C1D2}}{N_{D1C2}+N_{C1D2}}.
\end{equation}

For our experiment, we find $V_{\text{feed}}=0.56\pm0.06$, which
is identical to $V_{\phi}$ and equally within error bars of $V_{\text{ref}}$. We therefore conclude that the feedback works as expected, without introducing any loss of coherence.

\section{Conclusions}

We have demonstrated a technique for steering the measurement
of a quantum superposition towards a definitive result, a result which can be interpreted as ensuring that Schrödinger's cat is always observed alive. This was achieved
by using a feedback mechanism that enforces either bosonic or fermionic
behaviour on interfering photons with long coherence lengths.

We emphasize that a classical interpretation of the described experiment
fails. Classically, one might expect the feedback control to be successful
only when the first detection corresponds to the photon in the delay
arm. Otherwise, if the first detection was of the photon in the short
arm, the phase change of the driving laser would have no effect
(as the photon under generation has already been detected). In a classical description, one would expect a random routing of the second photon
regardless of the phase switch, and a reduction of the feedback visibility
to zero, which is clearly not the case.

This result constitutes an elementary step towards introducing active
control into processes such as quantum random walks and optical networks
\citep{Peruzzo_2010}. Generalisations of the technique
demonstrated here are suitable candidates in photonic switchyards requiring multiple photon streams for studying multi-mode
interferometry \citep{barrettMultimodeInterferometryEntangling2019},
where the deterministic routing of photons would be performed using feedback operations.

\section*{Acknowledgements}

J.R.A. acknowledges Alejandra Valencia, David Guzm\'{a}n and Sebasti\'{a}n
Murgueitio Ram\'{\i}rez for useful discussions.\\

We dedicate this paper to the memory of Bruce W. Shore, who sadly
passed away on 9 January 2021. Bruce inspired us to keep questioning
the fundamental principles of the light-matter interaction, and without
his deep insight, laid out convincingly in his memoir on our changing
views of photons \citep{shoreOurChangingViews2020}, we would never
have accomplished the present work.

\section*{Funding}

European Union Horizon 2020 (Marie Sklodowska-Curie 765075-LIMQUET).
\\
 EPSRC through the quantum technologies programme (NQIT hub, EP/M013243/1){\footnotesize{}.}{\footnotesize\par}

{\footnotesize{}\bibliographystyle{apsrev4-2}
\bibliography{MainText}
}{\footnotesize\par}
\end{document}


\global\long\def\ket#1{\left|#1\right\rangle }%

\global\long\def\bra#1{\left\langle #1\right|}%

\global\long\def\n{\frac{1}{\sqrt{2}}}%

\global\long\def\mye{\text{e}}%

\global\long\def\myi{\text{i}}%

\global\long\def\creatop#1{\hat{a}_{#1}^{\dagger}}%

\global\long\def\anniop#1{\hat{a}_{#1}}%

\title{Supplementary material:\\
How to administer an antidote to Schr\"{o}dinger's cat}
\maketitle

\section{Same-detector probabilities}

\begin{figure*}
\begin{centering}
\includegraphics[width=2\columnwidth]{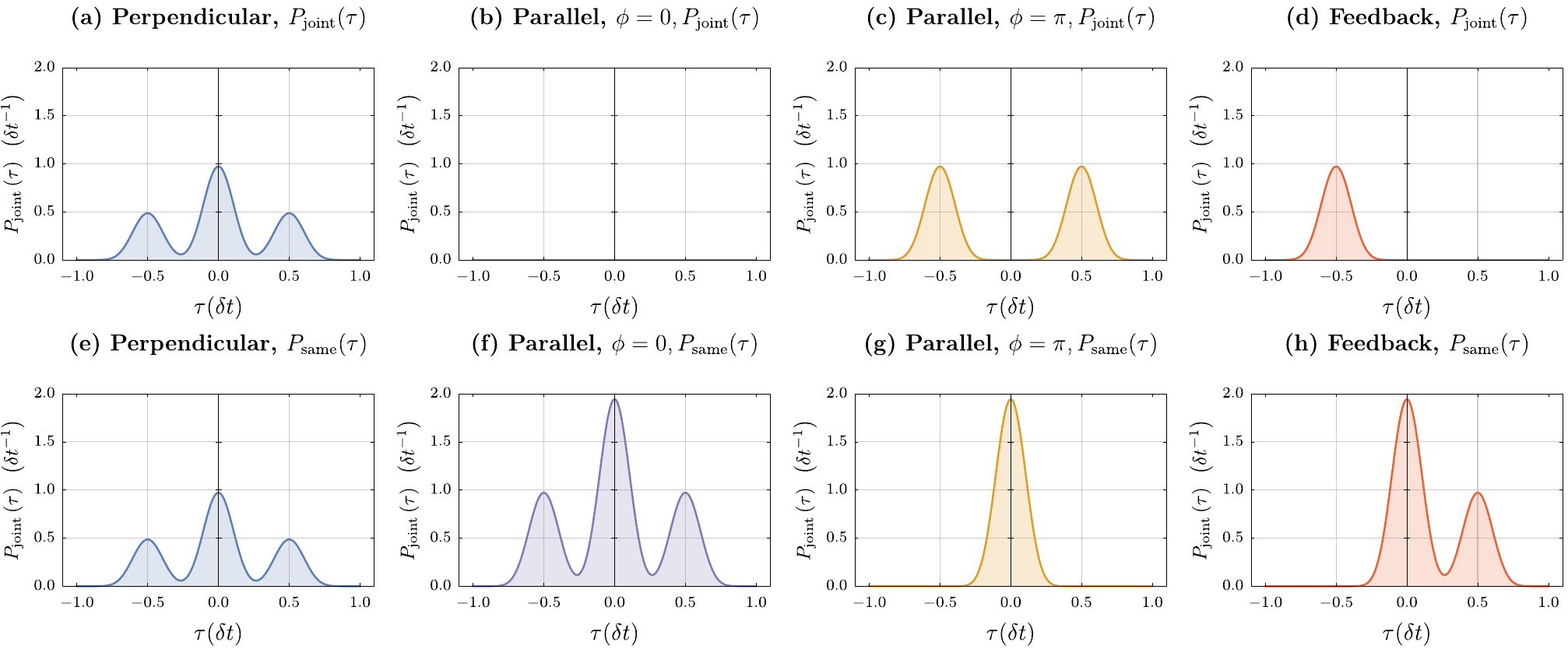} 
\par\end{centering}
\caption{Theoretical curves for $P_{\text{joint}} (\tau)$ for cross-detector (a-d) and same-detector coincidences (e-h), for the same cases described in Fig. 3 from the paper. Subfigure (f) shows that, when photons are identical, all the detections happen in the same detector, albeit possibly in different time bins. Once more, $\delta t$ corresponds to the length of the single photons.  
\label{fig:SameDetectorCrossDetector} }
\end{figure*}

Fig. 3 in the main text shows the subset of probabilities when we only look at cross-detector correlations. For this reason, the value $\int P_{\text{joint}}\left(\tau\right)d\tau$ is bounded by $1/2$. A prediction of the theoretical curves for same-detector correlations, given by $P_{\text{same}}\left(\tau\right)$ can be shown in Fig. \ref{fig:SameDetectorCrossDetector}. Given the different possibilities that can arise from photon routing, the normalization for the joint detection probability is given by

\begin{equation}
    \int\left(P_{\text{joint}}\left(\tau\right)+P_{\text{same}}\left(\tau\right)\right)d\tau=1
\end{equation}

\section{Latency of feedback}

To ensure adequate feedback, it is crucial to perform a timely change
on the phase $\phi$, conditional on detections only in the first
time bin. However, not every detection of the SPCM can be used to
change the phase of the photon: the latency between the SPCM click
and the effect of the feedback electronics can affect the production
of a proper photonic state which can be used for feedback. The photon signal takes around $75\units{ns}$ to reach the output of the SPCMs from the EOM (as seen in the sum of the first two columns of Table \ref{tab:LatencyBreakdown}), during which there is no possibility of achieving feedback control.

The probability of a correlation (with one detection in each time
bin) occurring during the dead time is given by 
\begin{equation}
P(\tilde{t})=\int_{\max\left[\frac{1}{2}-\tilde{t},0\right]\delta t}^{\frac{1}{2}}f\left(t_{1}\right)\int_{\frac{1}{2}}^{\min\left[t_{1}+\tilde{t},1\right]\delta t}f\left(t_{2}\right)\mathrm{d}t_{2}\mathrm{~d}t_{1}\label{eq:ProbOfCorr-1}
\end{equation}
 where $\tilde{t}$ is the dead time given as a fraction of the total photon
length $\delta t$, $t_{1}\in\left(0,\frac{1}{2}\right)\delta t$ and $t_{2}\in\left(\frac{1}{2},1\right) \delta t,$
and
\begin{equation}
f(t)=\frac{16}{3}\sin^{4}(2\pi t / \delta t)\label{eq:PhotonShape-1}
\end{equation}
 is the intensity envelope of the produced photons.

\begin{figure}
\begin{centering}
\includegraphics[width=0.9\columnwidth]{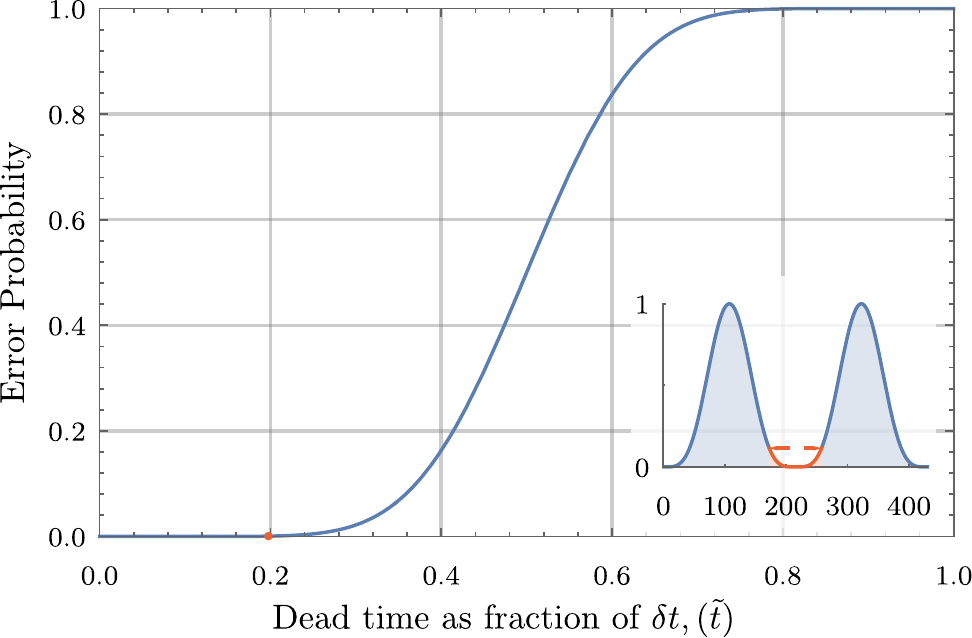}
\par\end{centering}
\caption{Feedback error analysis: The dead time is the feedback delay
of the system, here shown as a fraction of the photon length. Specifically
it is the round trip time for a signal to pass from the EOM, via the
cavity to the SPCMs, be processed by the controller and return to
the EOM. It represents a period of time during which feedback control
cannot be achieved, therefore introducing an error rate (Eq. \ref{eq:ProbOfCorr-1}).
The profile of the considered photon is shown in the inset along with
the measured $97\mathrm{~ns}$ feedback delay of the setup marked
in red, the corresponding error rate of $0.2\%$ is coincident with
the horizontal axis on this scale. \label{fig:Feedback-error-analysis-1}}
\end{figure}

The error rate remains negligibly small for dead times of up to around
a fifth of the total photon length before increasing rapidly, as shown
in Fig. \ref{fig:Feedback-error-analysis-1}. In our experiment, the
control and effecting stages are implemented in $13\text{ns}$, resulting in
a total latency for the feedback of $97\text{ns}$, corresponding to an expected
error rate of 0.2\%. Table \ref{tab:LatencyBreakdown} details the
latency breakdown.

\begin{table}
\begin{centering}
\begin{tabular}{|c|c|}
\hline 
Element & Time (ns)\tabularnewline
\hline 
Optical Transit Time $\mathrm{EOM}\rightarrow$ Cavity $\rightarrow$
SPCM & 45\tabularnewline
\hline 
SPCM Response & 35\tabularnewline
\hline 
Circuit Response & 7.0\tabularnewline
\hline 
Signal Rise Time & 5.5\tabularnewline
\hline 
Total Cable Delay $\mathrm{SPCM}\rightarrow$ Control $\rightarrow\mathrm{EOM}$ & 4.5\tabularnewline
\hline 
\hline 
Total & 97.0\tabularnewline
\hline 
\end{tabular}
\par\end{centering}
\caption{A breakdown of the contributing elements to the feedback delay. The
first two rows are not known absolutely, only their sum was measured.
From discussions with the manufacturer it is believed that the SPCM
response (time delay between photon impact and TTL output) is around
$35\units{ns}$.\label{tab:LatencyBreakdown}}
\end{table}

\section{Fast electronics}

\begin{figure}
\begin{centering}
\includegraphics[width=0.9\columnwidth]{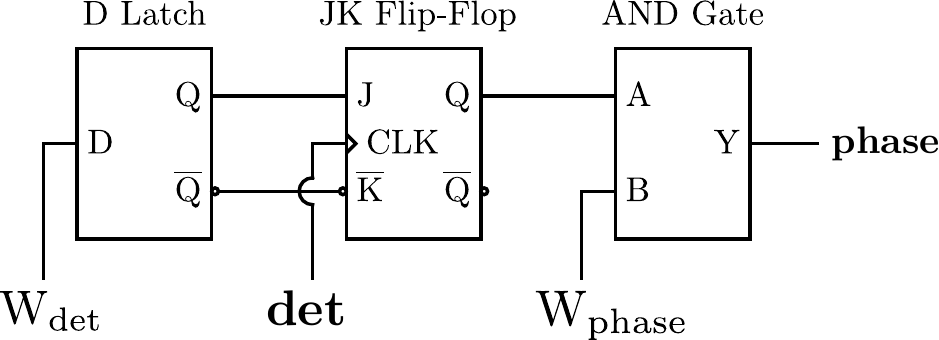}
\par\end{centering}
\caption{Control circuit: The JK flip-flop (TI SN74F109N) controls the state
of operation of the circuit. When $\mathrm{Q}$ is $0,$ no $\text{Phase}$
signal will be output, but when $\mathrm{Q}$ is $1,$ the output,
$\text{phase}$, follows $\text{W}_{{\text{phase}}}$,
using a wired AND gate with open collector TTL gates (N74F07N). Toggling
between these two states is triggered by a signal on the JK flip-flop
CLK (clock) input from a detector, $\text{det}$, but only when
the JK is in toggle mode. With $\mathrm{Q}(\overline{\mathrm{Q}})$
high (low) the JK is in toggle mode, with $\mathrm{Q}$ ($\overline{\mathrm{Q}}$
) low (high) the JK is in Hold mode. The D latch (TI SN74LS375N) buffers
the input $\textsc{W}_{\text{det}}$ to provide a duplicated and
negated copy necessary for the JK. \label{fig:FastElectronics}}
\end{figure}

The task of the feedback controller is to decide whether or not a
phase should be applied to the second time bin and, if so, quickly
supply sufficient voltage to the electro-optic modulator to enact
that phase change. To that end, a custom in-house circuit was built
using off the shelf transistor-transistor logic (TTL) integrated circuit (IC) logic chips. The circuit toggles between
one mode of operation with no effect and another one which ensures
a phase change to the EOM during the second time bin. An abbreviated
circuit logic diagram outlining the functioning of the circuit is
shown in Figure \ref{fig:FastElectronics}. The control circuit receives
as inputs a copy of the TTL output from the SPCM of interest, $\textbf{det}$, and two
$215\units{ns}$ TTL window pulses from the AWG, $\text{W}_{\text{det}}$
and $\text{W}_{\text{phase}}$, which respectively outline the
first time bin for the registering of detections and the second time
bin for the output phase voltage, $\textbf{phase}$. 

\section{Data processing}

The data recorded by the SPCMs is processed to yield the sliding histograms
and bar charts shown in Figures 3 and 4. This data contains a significant
amount of noise coming from detector dark counts and other stray photons,
which need to be corrected for. Five procedures are performed in the
raw data after clicks are recorded in detectors $C$ and $D$:\\

\begin{figure}
\begin{centering}
\includegraphics[width=0.8\columnwidth]{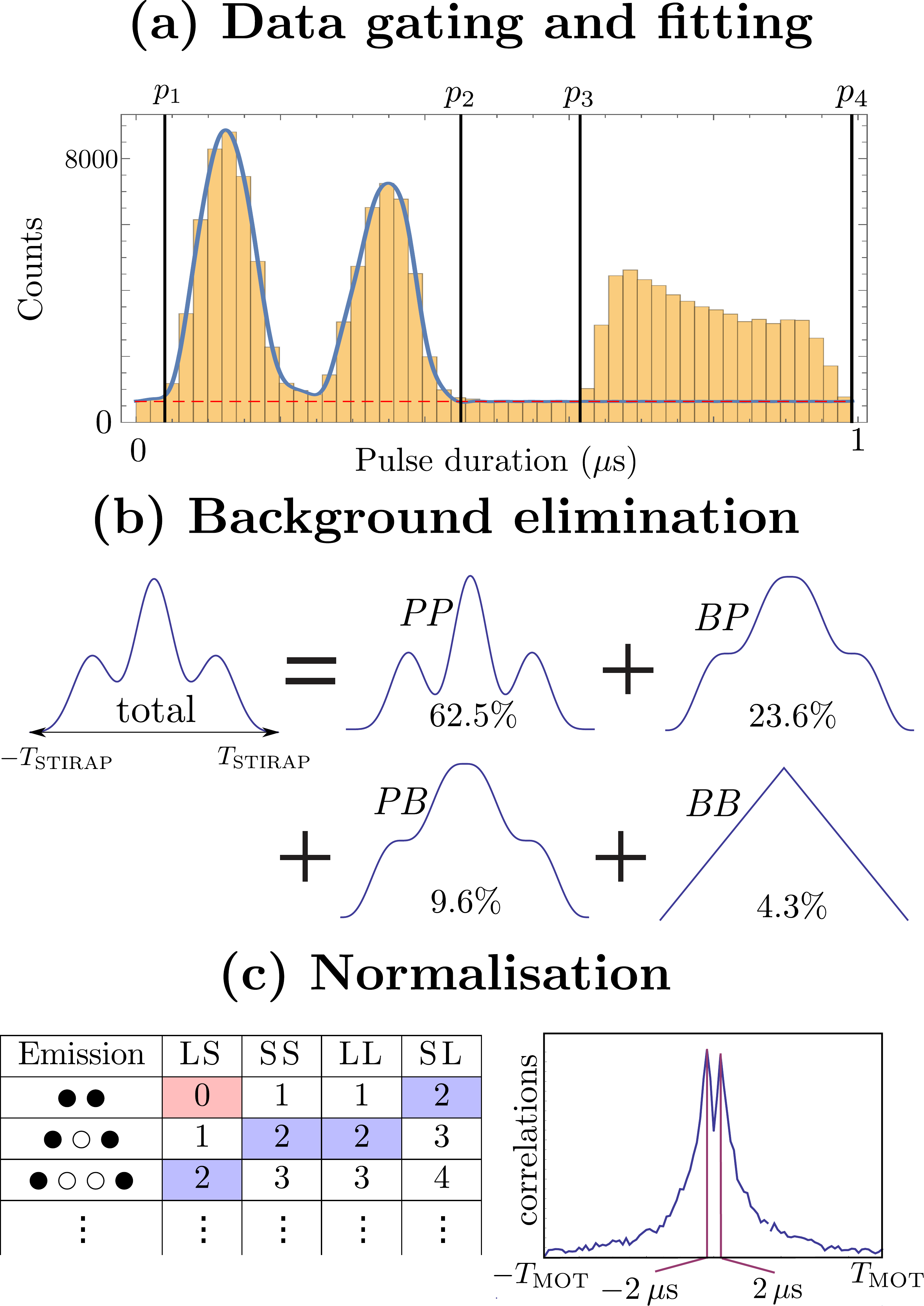}
\par\end{centering}
\caption{Background corrections. (a) Shows the photon gating, which fits the photon shape and the repumping following Eq. \ref{eq:Fitting}. The photon duration begins in $p_1$ and ends in $p_2$. (b) Shows the four components of the background elimination, $M_{\mathrm{BB}}^{(C)(D)}(\tau),M_{\mathrm{BP}}^{(C)(D)}(\tau)$,
$M_{\mathrm{PB}}^{(C)(D)}(\tau)\text{ and }M_{\mathrm{PP}}^{(C)(D)}(\tau)$. (c) shows the normalisation of the correlations. Coincident detections (red), can only occur for a sequential emission of two photons with the first photon entering the delay arm, L and the second entering the short arm S. The normalisation constant is based on the number of correlations at $\pm 2\units{\mu s}$ in the $g^{(2)}$ (blue), for which there are four times as many pathways. Correcting for photon losses in the delay arm, we find that the normalisation factor follows Eq. \ref{eq:normalisation}. $T_{\text{MOT}}$ corresponds to the time taken by an atom to be launched from the atomic fountain and fall back again, and the loss in correlations is a direct consequence of the passage of atoms through the cavity. \label{fig:DataProcessing}}
\end{figure}

\textbf{A. Gating the raw data:} Correlations need to be found between the raw data. The time scale
is cropped to account for photon arrival times and repumping times,
which are eliminated accordingly from the scale. This is shown in
Fig. \ref{fig:DataProcessing} (a). \\

\textbf{B. Fitting and calculation of Signal to Noise Ratio:} Once the raw data has been gated, we fit the following function to
the raw data in each detector: 
\begin{equation}
g\left(t\right)=a+\begin{cases}
b\sin^{2}\left(\frac{2 \pi \left(x-p_{1}\right)}{p_{2}-p_{1}}\right) & \text{for }p_{1}<x<p_{2}\\
c & \text{for }p_{3}<x<p_{4}
\end{cases}
\label{eq:Fitting}
\end{equation}

Here, $p_{1}$ and $p_{2}$ encompass the times for photon arrivals,
while $p_{3}$ and $p_{4}$ correspond to times when the repumping
is triggered. The time between $p_{2}$ and $p_{3}$ is used to obtain
the dark count rates of the single photon counting modules. This fitting
is shown in Fig. \ref{fig:DataProcessing} (b).\\

\textbf{C. Background estimation: } A model for calculating the background is created by using the photon
count and dark count rates of the single photon detectors. These rates
lead to two distributions for each detector describing the detection
events due to photons emitted from atoms, $m_{P}^{\left(C/D\right)}\left(t\right)$,
and true background detections uncorrelated to atoms $m_{B}^{\left(C/D\right)}\left(t\right).$

These distributions can be combined to give expected correlations
between their pairwise combinations across both detectors: $M_{\mathrm{BB}}^{(C)(D)}(\tau),M_{\mathrm{BP}}^{(C)(D)}(\tau)$,
$M_{\mathrm{PB}}^{(C)(D)}(\tau)\text{ and }M_{\mathrm{PP}}^{(C)(D)}(\tau)$,
where 
\begin{align}
\begin{array}{l}
M_{\mathrm{BP}}^{(C)(D)}(\tau)\end{array} & =\int_{0}^{T_{\mathrm{STIRAP}}}m_{\mathrm{B}}^{(C)}(t)m_{\mathrm{P}}^{(D)}(t+\tau)\mathrm{d}t\\
 & =\left(m_{\mathrm{B}}^{(C)}*m_{\mathrm{P}}^{(D)}\right)(\tau).
\end{align}
 and the other terms are defined analogously. The total background
is the addition of these terms:

\begin{multline}
M_{\text{total }}^{(C)(D)}(\tau)=M_{\mathrm{BB}}^{(C)(D)}(\tau)+M_{\mathrm{BP}}^{(C)(D)}(\tau)\\
+M_{\mathrm{PB}}^{(C)(D)}(\tau)+M_{\mathrm{PP}}^{(C)(D)}(\tau).
\end{multline}

The $M_{\mathrm{PP}}^{(C)(D)}$ term accounts for the expected correlation
rate between two atoms producing photons by the distribution $m_{P}^{\left(C/D\right)}$,
and can be discarded as we are only interested in the correlations
of photons within the same emission period. A plot of the four terms
composing $M_{\text{total }}^{(C)(D)}(\tau)$ is shown in Fig. \ref{fig:DataProcessing}
(c). Upon removing $M_{\text{total }}^{(C)(D)}(\tau)$, the ratio between the integrated coincidence rate and the integrated noise correction of the different contributions is given by Table \ref{tab:SNR}.\\

\begin{table}
\begin{centering}
\begin{tabular}{|c|c|}
\hline 
Case & Coincidence rate / noise \tabularnewline
\hline 
\hline 
Fig. 3(a) - Perpendicular & $2.02$\tabularnewline
\hline 
Fig. 3(b) - Parallel, $\phi=0$ & 0.37\tabularnewline
\hline 
Fig. 3(c) - Parallel, $\phi=\pi$ & 1.40\tabularnewline
\hline 
Fig. 3(d) - Parallel, feedback $\phi$ & 0.94\tabularnewline
\hline 
\end{tabular}
\par\end{centering}
\caption{Integrated coincidence rate and integrated noise correction ratio of the different time-resolved HOM histograms.
The noise has been eliminated from Fig. 3 in the main text.}
\label{tab:SNR}
\end{table}

\textbf{D. Maximum likelihood estimation of correlation counts: }
The total number of observed correlation counts $O(\tau,\delta\tau)$ between $\tau$ and $\tau+\delta\tau$ is equal to
\begin{equation}
  \begin{aligned}
O(\tau, \delta\tau)=S(\tau,\delta\tau)+B(\tau, \delta\tau),
\end{aligned}  
\end{equation}
i.e. the sum of signal correlation counts $S(\tau,\delta\tau)$ and background correlation counts $B(\tau, \delta\tau)$ calculated from $M_{\text{total }}^{(C)(D)}(\tau)$. From $O(\tau,\delta\tau)$, we can extract the most likely mean number of signal correlation counts within such a bin with a maximum likelihood estimation. Assuming that both $S(\tau, \delta\tau)$ and $B(\tau, \delta\tau)$ follow independent Poisson distributions with parameters $\lambda_S$ and $\lambda_B$, respectively, it can be shown that the probability of observing $n$ counts equals
\begin{equation}
  \begin{aligned}
\operatorname{P}(O=n) 
&=\frac{e^{-(\lambda_S+\lambda_B)}(\lambda_S+\lambda_B)^{n}}{n !},
\end{aligned}  
\end{equation}
i.e. $O(\tau, \delta\tau)$ follows a Poisson distribution with a mean of $\lambda_O=\lambda_S+\lambda_B$. Given an observation $n$ for $O$ and known $\lambda_B$ (mean background), the likelihood function for the parameter $\lambda_S\in[0,\infty)$ is given by
\begin{equation}
\mathcal{L}\left(\lambda_S\mid n, \lambda_B\right)= \operatorname{P}_{\lambda_S}(O=n),
\end{equation}
which is maximised for the intuitive value of
\begin{equation}
\lambda_S = \max\{0, n-\lambda_B \},
\end{equation}
i.e. the number of observed counts minus the background, constrained to a positive number.\\

\textbf{E. Normalisation: } Due to experimental errors in detectors, such as losses and dead times, it is not possible to identify directly how many individual photon-pair experiments were performed by counting the number of detection events measured in one detector. To overcome this, it is possible to notice that the number of events happening simultaneous is related to the number of events happening separated by two duty cycles, following the relation 
\begin{equation}
N_{0}/N_{2}=\frac{\eta_{L}}{1+2\eta_{L}+\eta_{L}^{2}}\simeq1/4,
\label{eq:normalisation}
\end{equation}
where $\eta_L$ is the probability of photon transmission in the fibre delay.